# Transport Information System using Query Centric Cyber Physical Systems (QCPS)

| Ankit Mundra | Geetanjali Rathee | Meenu Chawla | Nitin Rakesh | Ashutosh Soni |
|---|---|---|---|---|
| Department of Computer Science and Engineering, Jaypee University of Information Technology, Waknaghat Solan, India-173234 | Department of Computer Science and Engineering, Jaypee University of Information Technology, Waknaghat Solan, India-173234 | Department of Computer Science and Engineering, Jaypee University of Information Technology, Waknaghat Solan, India-173234 | Department of Computer Science and Engineering, Jaypee University of Information Technology, Waknaghat Solan, India-173234 | Department of Computer Science and Engineering, Jaypee University of Information Technology, Waknaghat Solan, India-173234 |

## ABSTRACT
To incorporate the computation and communication with the physical world, next generation architecture i.e. CPS is viewed as a new technology. To improve the better interaction with the physical world or to perk up the electricity delivery usage, various CPS based approaches have been introduced. Recently several GPS equipped smart phones and sensor based frameworks have been proposed which provide various services i.e. environment estimation, road safety improvement but encounter certain limitations like elevated energy consumption and high computation cost. To meet the high reliability and safety requirements, this paper introduces a novel approach based on QCPS model which provides several users services (discussed in this paper). Further, this paper proposed a Transport Information System *(TIS)*, which provide the communication with lower cost overhead by arranging the similar sensors in the form of grids. Each grid has a coordinator which interacts with cloud to process the user query. In order to evaluate the performance of proposed approach we have implemented a test bed of 16 wireless sensor nodes and have shown the performance in terms of computation and communication cost.

## 1. INTRODUCTION
Gradually, Internet technologies and researchers have stirred from read only web to read/write web and more recently towards intellectual web [1]. The next phase of internet development is in Cyber Physical System *(CPS)* that allows direct manipulation of physical world through the internet. The latest pervasive deployment of Wireless sensor network *(WSN)* into *CPS* has brought vast amount of data about the physical world into cyber world as sensors [1] (which determine the state of real world objects). To provide better interaction with physical world, we need to have more effective communication and computation mechanism for multi-heterogeneous network. Today *CPS* is viewed as a new technology for future engineered [2], which is a next generation architecture that incorporates computation and communication with the physical world. With this *CPS* has been recognized as a key technology which provides next generation applications that improves transportation management, electricity usage, delivery and health-care [1-3].

In Transportation Information Systems *(TIS)*, *CPS* aims to integrate computing/communication capabilities that support various applications, for example: road safety improvement, velocity and travel time estimation, environment estimation, on-road infotainment etc [4-5]. With this, the *TIS* also give consideration to the traffic congestion and delay issues because it spreads inefficiency, user frustration and wasted fuel [6].

Over past few years, several *GPS* equipped smart phones and sensor based models were proposed as traffic probes, which still encounter certain limitations i.e. elevated energy consumption, high computation and communication cost. To overcome from these drawbacks Query centric cyber physical system *(QCPS)* model becomes very useful. *QCPS* model based on cyber physical system which enables the fast execution of centric query (i.e. a single query which incorporates the multiple heterogeneous sensor nodes) with less computation and communication cost and supports low energy consumption [7-8]. *QCPS* architecture (Fig. 1) consists of three interdependent things 1) group of networks which sensed the data and send it to the cloud for computation. 2) Cloud that maintains the separate database for each type of network. 3) Group of users which enable the users to directly interact with cloud centric query processing [5].

In this paper *QCPS* based *TIS* have been introduced which integrates various computing/communication capabilities and provide a centric query result to user. *TIS* approach process the centric query of user by arranging the similar sensors in to a single grid further, sensors having the distance greater than threshold but of similar types are assigned to different grid. Each grid has coordinator which sensed the nodes of its grid. Whenever a user request for a query, coordinator communicate with cloud, process query and send it back to user. In *TIS* provides several services to user (i.e. road condition estimation, environment estimation, velocity and travel time estimation, traffic congestion information etc.) by centric query.

The rest of this paper is organized as follows. Section 2 presents previous approaches related to the work done in this research. Section 3 presents our proposed approach (*TIS*) based on previously proposed *QCPS* [8] that provides various services (defined above) according to user's centric query. Section 4 describes the performance evaluation and results of proposed system by implementing a testbed of 16 sensor nodes. Section 5 concludes the paper and discusses the future scope of work.

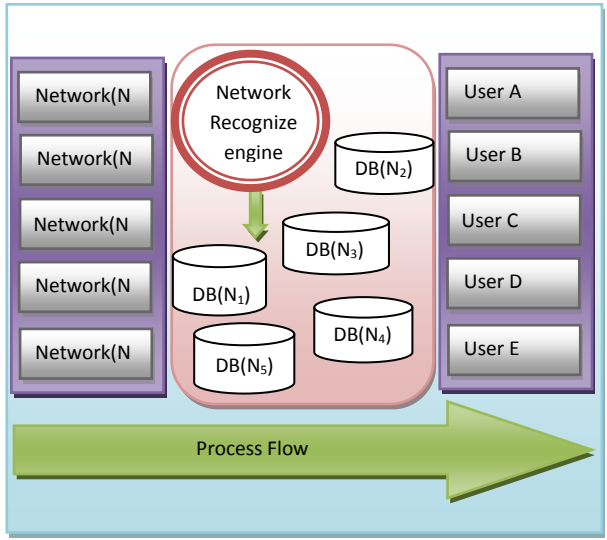

**Fig 1: QCPS Architecture [7]**

## 2. RELATED WORK
Cyber physical system has been recognized as a key technique for providing next generation application which improves electricity delivery and usage, health care, transportation management system [1]. To meet the high reliability and safety requirements for these systems, various GPS based applications, frameworks have been proposed. [9] Describe how the mobile internet is changing the face of transportation CPS. They built a traffic monitoring system known as mobile millennium by using GPS equipped mobile devices, in this system collect traffic data from GPS-equipped mobile phones and estimate traffic condition in real time. The major drawback of this approach is high battery consumption, each mobile phone should be GPS equipped. [1] Proposed a web-of-thing based CPS architecture to improve road safety and to achieve high demand response in smart grids. [10] Proposed a framework on SIP/ ZigBee architecture. In this by using SIP and its extension, a seamless convergence of traffic measurement and short-range wireless sensor and equator networks can be achieved. [11] Proposed traffic monitoring system based on vehicle based sensor networks. These vehicle based sensors are embedded in vehicle for monitoring the traffic. These sensors are used due to its low communication cost and avoid the network congestion. The major drawback of this system is that it does not define the relationship between accuracy and number of speed elements (SE). So, these proposed systems suffer with some drawbacks and does not provide the services efficiently. To provide these services like Environment estimation, Road Condition estimation, Velocity and Travel Time estimation and Traffic Congestion information etc, we propose novel *TIS* based on QCPS that provides services according to user's query.

## 3. PROPOSED APPROACH
In this section we have illustrate the proposed approach i.e. Traffic Information Systems *(TIS)* which incorporates the Query centric cyber physical system model [5] and integrate various computing/communication capabilities that provide a variety of services i.e. road condition estimation; environment estimation; velocity and travel time estimation; and traffic congestion information. The main goal of our approach is to achieve significant reduction in traffic fatalities and serious injuries on all public roads and highways. In order to provide these services efficiently various types of sensors are used (Table 1) that sense and estimate the conditions and give information to users at any time. 1) In Road Condition estimation various Vision sensors are used that sense and estimates the condition of road like the road is distorted or not? And whether the road is single lane and double lane? With this it estimates the traffic congestion on the road and gives reliable information to user for their safety. 2) The velocity and travel time estimates the speed of traffic and determine the distance between two traffics. 3) The environment estimation service assesses the condition of weather i.e. whether it is cloudy/smoky or not? And measure the temperature of weather thus for several environment sensors are used like temperature sensors, humidity sensors, light sensors etc.4) Miscellaneous Sensors are used to monitor vehicle crash, traffic congestion, etc.

**Table 1. Table captions should be placed above the table**

| Types of Sensors | Purpose |
| --- | --- |
| **Vision Sensors** | road safety improvement |
| **Speed Sensors** | velocity and travel time estimation |
| **Environment Sensors** | environment estimation |
| **Miscellaneous** | Vehicle crash, congestion etc. |

Further, Fig. 2 shows the working illustration of *TIS* i.e. each type of sensor sense the data respectively (discussed earlier). As shown in below figure sensor node embedded at tower 1 is falls under Miscellaneous category (according to our classification) and sense vehicle crash, traffic condition; sensor node situated on tower 2 sense road condition like merging of road, digs on road etc. Sensor node on tower 3 senses the traffic condition also type of vehicle whether there are two wheeler, heavy duty, trucks, buses etc. Further sensor on tower 4 is used to sense the weather condition like rain, temperature, humidity etc. In order to support the query centric facility to user (by a single query user can get complete result) *TIS* adopt the mechanism of QCPS which is shown in Fig. 3.

### 3.1 Working of TIS
To understand the working process of *TIS* we have described following steps:

Step 1. The sensors which of same type and have distance less than the threshold distance are assigned to same grid.

Step 2. From each grid, a coordinator is elected and have responsible to communicate with all the sensor nodes of its grid and with cloud database also (coordinator is elected by mechanism described in [5]).

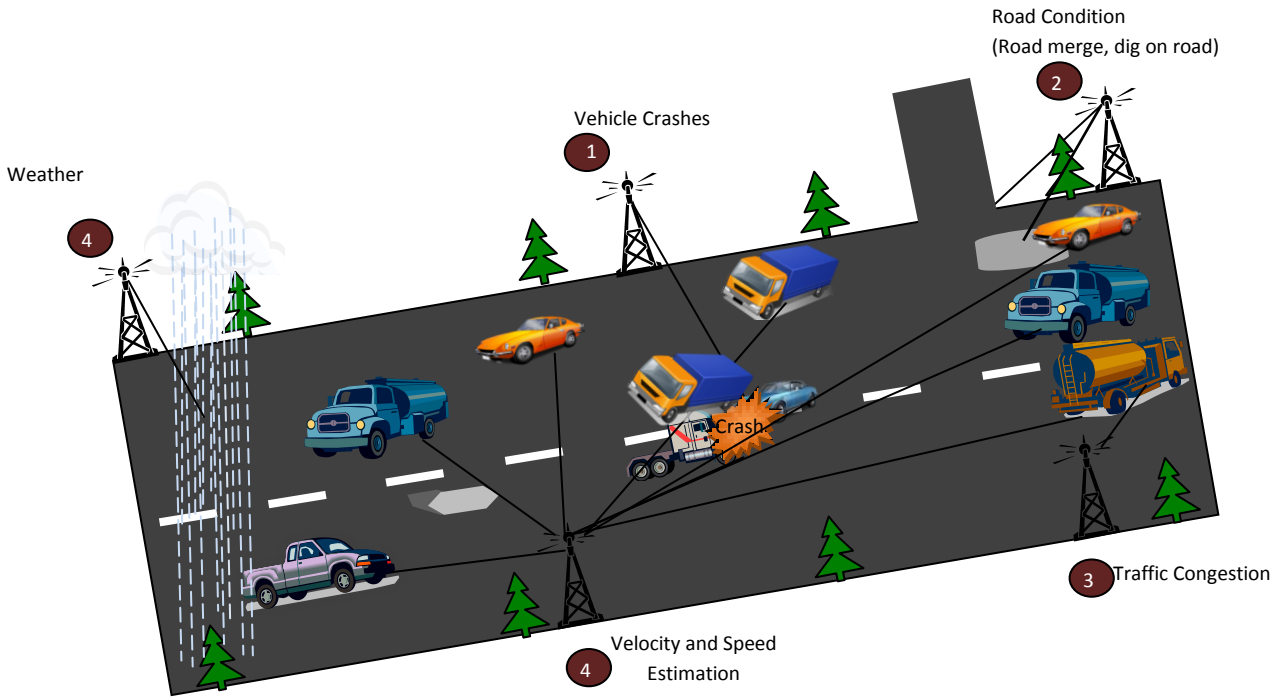

**Fig 2: An Illustration of TIS**

Step 3. Each coordinator node has a base station from where its respective grid is controlled.

Step 4. In cloud there is database created for each type of sensor nodes and have separate table in a database for each sensor node.

Step 5. Whenever a sensor node needs data of other sensor which is situated in its own grid or in some other grid it has to communicate with the coordinator sensor node of its own grid.

Step 6. Coordinator communicates with the database in cloud and fetches the desired result (needed computation is performed in cloud), than provide this result to demanding sensor node.

Step 7. When user generates a query, computation is performed in the databases stores in cloud. Then after user gets complete estimation report.

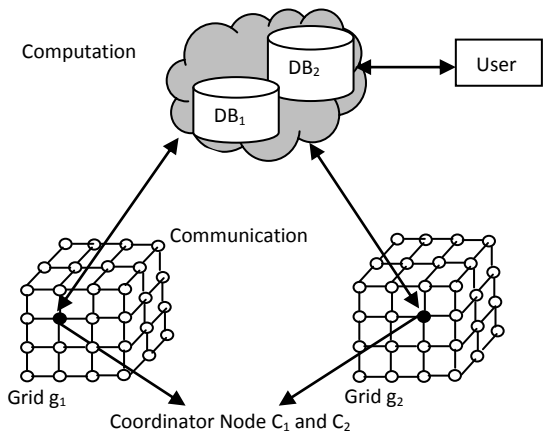

**Fig 3:** *QCPS Model* **[7]**

Further we have provided the TIS algorithm which is shown below:

## 3.2 TIS Algorithm

| **Step 1: Initial Condition:** Different Types of Sensors **Final Condition:** Return User Based Query i.e. Road Safety, Velocity and Travel Time Estimation, Environment estimation, on-road infotainment **Given:** Types of Sensors, Physical Location, Cloud, Base Station **Local variables:** Distance, Type, i, j, N |
|---|
| **Step 2:** Initialize Local Variables i, j, N with Zero <br>     For all node N=1, 2 ...n <br>         For i= 1 to n <br>             For j= i+1 to n <br>                 If <br>                 */\* It checks two conditions first is type of two sensor nodes and second is distance between two nodes should less than pre define distance\*/* <br>         Then <br>             They are assigning to a same grid; <br>     End if <br>     Else <br>         Repeat; <br>     End else |
| **Step 3:** Elect coordinator from each grid and allow communicating with cloud database; |
| **Step 4:** Process user query among ($c_1, c_2, c_3, c_4$); <br> **Step 5:** Return; |

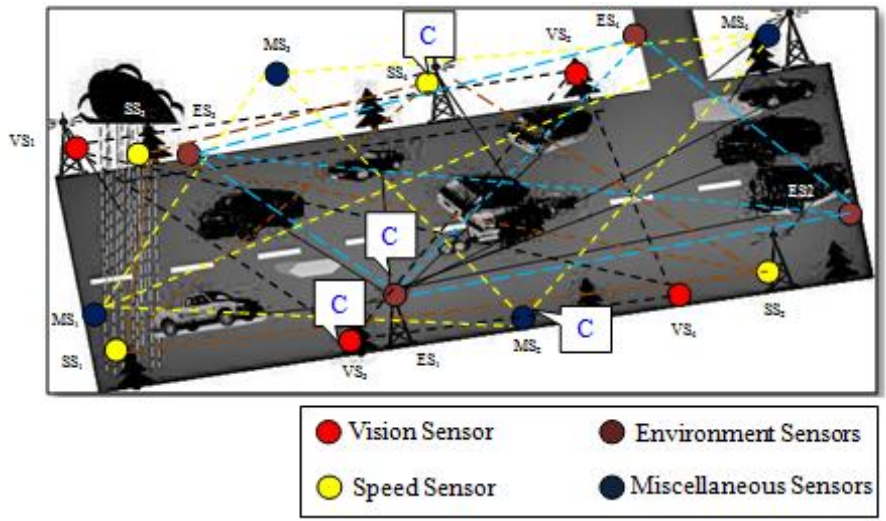

**Fig 4: Testbed of TIS**

## 4. PERFORMANCE EVALUATION

To evaluate the performance of *TIS* we have implemented a testbed which consist of sixteen sensor nodes (four sensor nodes of each type) and these sensors are deployed on the road near by our university (virtual scenario is created on Unix platform). Fig. 4 shows the formulation of testbed in which communication between same types of sensors is shown by dashed lines. Further, among the five sensors a node has been elected as coordinator node. Each coordinator has a base station from where the data of its grid is stored in cloud database (Table 2). In the below figure letter *C* shows that this is the coordinator sensor node.

**Table 2. Cloud Database of Sensors**

| Sr. No. | Sensor Type | Database Name |
|---|---|---|
| 1. | Vision | VDB |
| 2. | Speed | SDB |
| 3. | Environment | EDB |
| 4. | Miscellaneous | MDB |

Table 3. Shows location of the sensor nodes which are implemented in our testbed. Sensors which comes into Vision sensor category ($VS_1,VS_2,VS_3,VS_4$) store there data in VDB, similarly Speed sensors ($SS_1$, $SS_2$, $SS_3$, $SS_4$) store there data in SDB, Environment sensors ($ES_1$, $ES_2$, $ES_3$, $ES_4$) store there data in EDB and Miscellaneous sensors ($MS_1$, $MS_2$, $MS_3$, $MS_4$) store there data in MDB in cloud database. Here, Vision sensors form one grid i.e. $G_1$, Speed sensors form grid $G_2$, Environment sensors form grid $G_3$, whereas Miscellaneous sensors form grid $G_4$. Further, after the execution of testbed user gets a centric result which consists of the result or sensed data of each type of sensor node.

The performance of *TIS* is evaluated upon two parameters that are computation and communication cost. *TIS* provide efficient execution of centric query and take less communication and computation time and thus it also resolves the issue of energy in wireless sensor networks (shown in table 4).

**Table 3. Sensors Location**

| Type of Sensor | Name of Sensor | Coordinate |
|---|---|---|
| **Vision Sensor** | $VS_1$ | (5,45,48) |
| | $VS_2$ | (85,43,75) |
| | $VS_3$(C) | (38,35,12) |
| | $VS_4$ | (89,56,23) |
| **Speed Sensor** | $SS_1$ | (7,36,10) |
| | $SS_2$ | (94,47,80) |
| | $SS_3$ | (16,35,67) |
| | $SS_4$(C) | (42,29,63) |
| **Environment Sensor** | $ES_1$(C) | (37,41,15) |
| | $ES_2$ | (104,35,24) |
| | $ES_3$ | (33,48,56) |
| | $ES_4$ | (86,39,74) |
| **Miscellaneous Sensor** | $MS_1$ | (4,42,11) |
| | $MS_2$(C) | (58,37,23) |
| | $MS_3$ | (47,44,46) |
| | $MS_4$ | (99,38,75) |

**Table 4. Effect on Cost (communication and computation)**

| Communication Cost | Reduced | Because a sensor have to communicate with the nearby centroid sensor only, not with the node which is large distant apart. |
|---|---|---|
| Computation Cost | Reduced | All calculation is performed in cloud as a request arises |

## 5. CONCLUSION AND FUTURE WORK

In this paper we have proposed an approach *(TIS)* which reduces the traffic problem and provides traffic information to user very efficiently. Further, by implementing a testbed of 16 sensor nodes we have evaluate the performance of *TIS* approach which shows that it require less communication and computation power for centric query processing.

In future we will implement the *TIS approach* in real world with large sensor networks and will study the result in real time environment.

## 6. REFERENCES

[1] Singh, Jaipal, and Omar Hussain, 2012, Cyber-Physical Systems as an Enabler for Next Generation Applications, In Network-Based Information Systems (NBiS), 2012 15th IEEE International Conference on, pp. 417-422.

[2] Mundra Ankit, Bhagvan K. Gupta, Geetanjali Rathee, Meenu Chawla, Nitin Rakesh, and Vipin Tyagi, Validated Real Time Middle Ware For Distributed Cyber Physical Systems Using HMM, arXiv preprint arXiv:1304.3396 (2013).

[3] Jin Wang, Hassan Abid, Sungyoung Lee, Lei Shu, Feng Xia, 2011, A Secured Health Care Application Architecture for Cyber- Physical Systems, Vol. 13, No. 3, pp. 101-108.

[4] W. Jones, 2001, Forecasting traffic flow, IEEE Spectrum, Vol.38, No. 1, pp. 90–91.

[5] Li, Xu, Chunming Qiao, Yunfei Hou, Yunjie Zhao, Aditya Wagh, Adel Sadek, Liusheng Huang, and Hongli Xu, 2013, On-road ads delivery scheduling and bandwidth allocation in vehicular CPS, In INFOCOM, 2013 Proceedings IEEE, pp. 2571-2579.

[6] Thiagarajan, Arvind, Lenin Ravindranath, Katrina LaCurts, Samuel Madden, Hari Balakrishnan, Sivan Toledo, and Jakob Eriksson, 2009, VTrack: accurate, energy-aware road traffic delay estimation using mobile phones, In Proceedings of the 7th ACM Conference on Embedded Networked Sensor Systems, pp. 85-98. ACM, 2009.

[7] Mundra, Ankit, Nitin Rakesh, and Vipin Tyagi. "Query Centric CPS (QCPS) Approach for Multiple Heterogeneous Systems." *arXiv preprint arXiv:1306.6397* (2013).

[8] Yuan He, 2012, COSE: A Query-Centric Framework of Collaborative Heterogeneous Sensor Networks, IEEE transactions on parallel and distributed systems, Vol. 23, No. 9, pp. 1681-1693.

[9] Work, D., and A. Bayen, 2008, Impacts of the mobile internet on transportation cyberphysical systems: traffic monitoring using smartphones, In National Workshop for Research on High-Confidence Transportation Cyber-Physical Systems: Automotive, Aviation, & Rail, pp. 18-20.

[10] Zhou, Jianhe, Chungui Li, and Zengfang Zhang, 2011, Intelligent transportation system based on SIP/ZigBee architecture." In Image Analysis and Signal Processing (IASP), 2011 IEEE International Conference on, pp. 405-409.

[11] Li, Xu, Wei Shu, Ming-Lu Li, and Min-You Wu, 2008, Vehicle-based Sensor Networks for Traffic Monitoring." In Proc. of 17[th] IEEE International Conference on Computer Communications and Networks.